\documentclass{article}
\usepackage{spconf}
\usepackage{cite}
\usepackage[pdftex]{graphicx}
\graphicspath{{Figure/}}
\usepackage[cmex10]{amsmath}
\usepackage{amsthm}
\usepackage{amssymb}
\usepackage{algorithmic}
\usepackage{array}
\usepackage{color}
\usepackage{epstopdf}
\usepackage{amsfonts}
\usepackage{gensymb}
\usepackage{hyperref}

\usepackage{pgfplots}
\pgfplotsset{compat=1.3}
\usepackage{tikz}							
\usetikzlibrary{shapes}
\usetikzlibrary{spy}
\usetikzlibrary{circuits}
\usetikzlibrary{arrows}

\newcommand{\vect}[1]{\boldsymbol{\mathrm{#1}}}

\title{WaveComBox: a Matlab Toolbox for Communications\\ using New Waveforms}

\name{F. Rottenberg$^{\dagger \star}$ \quad M. Van Eeckhaute$^{\star}$ \quad T.-H. Nguyen$^{\star}$ \quad F. Horlin$^{\star}$ \quad J. Louveaux$^{\dagger}$\thanks{The research reported herein was partly funded by the Fonds de la Recherche Scientifique (FNRS). The authors also want to thank Nicolas Boumal and Pierre-Yves Gousenbourger for their precious advice.} 
}

\address{$^{\dagger}$Universit\'e catholique de Louvain, 1348 Louvain-la-Neuve, Belgium \\
	$^{\star}$Universit\'e libre de Bruxelles, 1050 Brussel, Belgium }

\begin{document}
%
\maketitle
\begin{abstract}
Future generations of communications systems will target a very wide range of applications. Each application comes with its own set of requirements in terms of data rate, latency, user density, reliability... Accommodating this large variety of specifications implies the need for a novel physical layer technology. In this regard, the most popular modulation nowadays, namely, the orthogonal frequency division multiplexing modulation, is characterized by a poor time-frequency localization, implying strong limitations. In the light of these limitations, communications using new waveforms, relying on more sophisticated signal processing techniques and providing improved time-frequency localization, have attracted a lot of attention for the last decade. At the same time, the higher complexity of these new waveforms, not only in terms of implementation but also conceptually, creates an entrance barrier that slows down their adoption by industries, standardization bodies and more generally in the telecommunication community. The WaveComBox toolbox, freely available at \href{www.wavecombox.com}{\texttt{www.wavecombox.com}}, is a user-friendly, open-source and well documented piece of software aiming at considerably lowering the entrance barrier of recently proposed waveforms. This article first describes the general abstract structure of the toolbox. Secondly, examples are given to illustrate how to use the toolbox and to show some more advanced functionalities.
\end{abstract}
\section{Introduction}
\label{sec:intro}

The standards for the future generations of communication systems let us expect revolutionary changes in terms of data rate, latency, energy efficiency, massive connectivity and network reliability \cite{shafi20175g,wu2017overview}. The network should not only provide very high data rates but also be highly flexible to accommodate a considerable amount of devices with very different specifications and corresponding to different applications, such as the Internet of Things, the Tactile Internet or vehicle-to-vehicle communications. These high requirements will only be met by introducing innovative technologies radically different from existing ones.

\begin{figure}[!t]
	\centering
	\resizebox{0.5\textwidth}{!}{%
		\Large
		\includegraphics[scale=1]{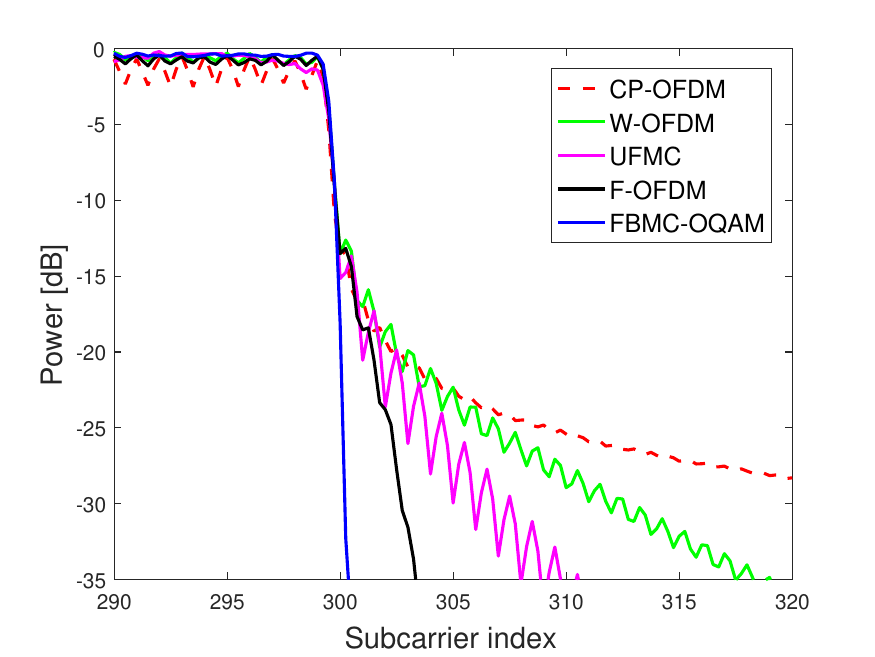}
	}
	\caption{Power spectral density (PSD) of different waveforms.}
	\label{fig:PSD}
\end{figure}

The orthogonal frequency division multiplexing modulation (OFDM) modulation is the most popular multicarrier modulation scheme nowadays. The main advantage of OFDM is its simplicity. Thanks to the combination of the Fast Fourier transform (FFT) and the introduction at the transmitter of redundant symbols known as the cyclic prefix (CP), the OFDM modulation allows for a very simple compensation of the channel impairments at the receiver \cite{li2006orthogonal}. However, the rectangular pulse shaping of the FFT filters induces significant spectral leakage, which results in the need for large guard bands at the edges of the spectrum in order to prevent out-of-band emissions (see Fig.~\ref{fig:PSD}). This bad frequency localization decreases the system flexibility regarding spectrum allocation and makes it less suited for applications such as cognitive radios and the Internet of Things, which may require asynchronous transmission for multiple users.

These limitations may be very detrimental for future generations of communications systems where the modulation format should at the same time be highly flexible and achieve high spectral efficiency. In this sense, a good time-frequency localization is very desirable. This has motivated research for new waveforms that would better fit these requirements \cite{Sahin2014}. This research has been conducted in parallel in many fields of communications including wireless communications \cite{6923528}, optical fiber communications \cite{horlin2013dual,7932847}, fiber-wireless communications \cite{Rottenberg18,rottenberg2018fbmc} or visible light communications \cite{lin2016experimental}. Actually, the research regarding waveform design has a long history and dates back to the sixties. This area of research has regained a lot of attention recently and a very large number of new waveforms have flourished, each one having its own specificity \cite{van2017performance,gerzaguet20175g}.

We now briefly discuss some of the main waveform contenders that are currently implemented in the toolbox. Their corresponding power spectral density (PSD) is shown in Fig.~\ref{fig:PSD}. To improve the frequency localization of the CP-OFDM modulation, the weighted overlap and add based OFDM (W-OFDM)  modulation uses a window to filter the square pulse used by the CP-OFDM modulation around each subcarrier, which leads to better spectral properties \cite{Zayani2016}. The FBMC-OQAM modulation uses purely real symbols (instead of complex symbols) at twice the symbol rate, resulting in a maximal spectral efficiency and a very good time-frequency localization. Demodulation is made easier by ensuring that the prototype filter satisfy real orthogonality condition \cite{farhang2011ofdm}. Instead of using a more refined filtering process at the subcarrier level, many schemes have been proposed recently, which perform an improved filtering at the resource block level, \textit{i.e.}, on a group of subcarriers: namely,  Universal filtered multicarrier (UFMC)  \cite{Vakilian2013} 
and filtered-OFDM (F-OFDM) \cite{Abdoli2015}. This type of systems has the advantage of keep a relatively a high compatibility with current OFDM systems. 

The advantages of these new waveforms generally come at the price of an increased complexity which, we believe, has slowed down their adoption by the community. This increased complexity does not only come from the more complex hardware architecture of the modulator and demodulator. More importantly, the new waveforms are conceptually more complex to apprehend and to implement. They require a deep re-thinking of the whole communication chain, implying the adaptation of general algorithms used for conventional signal processing operations such as channel estimation or equalization. The aim of the WaveComBox toolbox is to lower the entrance barrier of the new waveforms by allowing simple implementation of their physical layer functionalities.

By using an abstract architecture, the toolbox is made user-friendly, easy to apprehend and flexible. It addresses both single-input-single-output (SISO) and multiple-input-multiple-output (MIMO) configurations and implements conventional physical layer signal processing operations such as modulation and demodulation, channel estimation, channel equalization, synchronization... The channel models included in the toolbox may represent impairments typical from wireless and optical fiber mediums. The toolbox is open-source, allowing for easily checking and modifying the source code. It is documented with help files and examples. Finally, a forum is available to help users discuss of their problem and propose new contributions to the toolbox.

\begin{figure*}[t!]
	\centering
	\resizebox{0.9\textwidth}{!}{%
		{\includegraphics[clip, trim=0cm 12cm 5cm 0cm, scale=1]{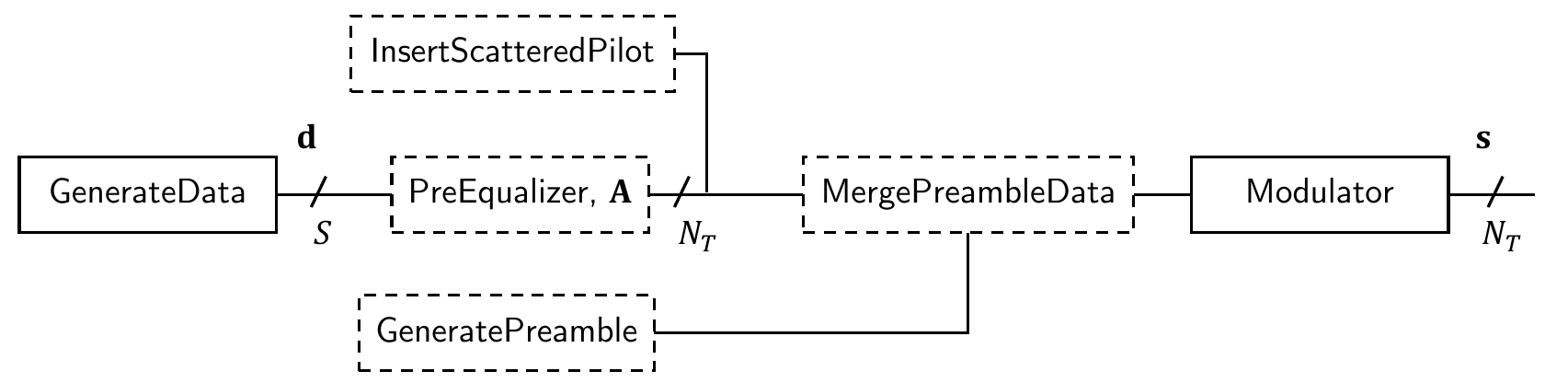}}  
	}
	\caption{Transmitter abstract block diagram.}
	\label{fig:transmitter}
\end{figure*}

\section{General architecture}
\label{sec:archi}

\begin{figure*}[t!]
	\centering
	\resizebox{0.4\textwidth}{!}{%
		{\includegraphics[clip, trim=0cm 16cm 23cm 0cm, scale=1]{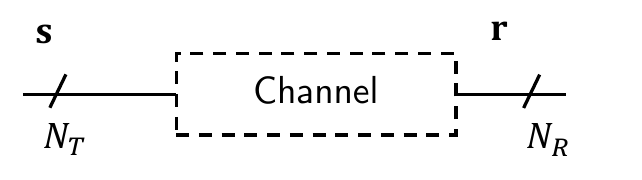}}  
	}
	\vspace{-1em}
	\caption{Channel abstract block diagram. Dashed boxes are optional.}
	\label{fig:channel}
\end{figure*}

\begin{figure*}[t!]
	\centering
	\resizebox{0.9\textwidth}{!}{%
		{\includegraphics[clip, trim=0cm 12cm 7cm 0cm, scale=1]{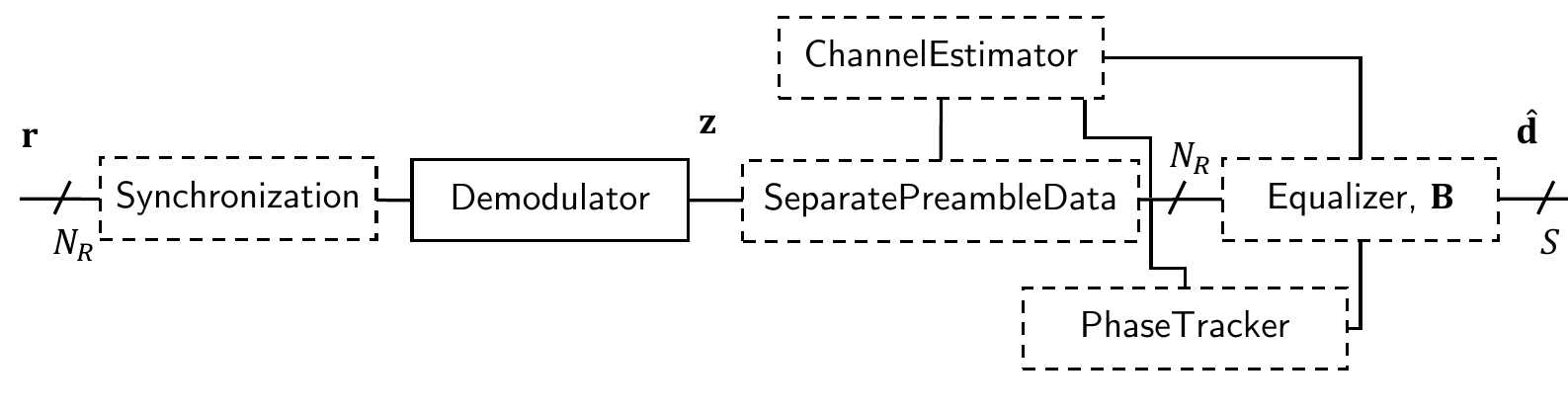}}  
	}
	\vspace{-1em}
	\caption{Receiver abstract block diagram. Dashed boxes are optional.}
	\label{fig:receiver}
\end{figure*}

The WaveComBox toolbox aims at implementing a complete communication chain relying on a specific waveform. The general architecture of the toolbox is divided in three mains parts: transmitter, channel and receiver. These three parts are depicted in the three block diagrams of Figs.~\ref{fig:transmitter}, \ref{fig:channel} and \ref{fig:receiver}. Each box consists of a basic signal processing block and corresponds to a function implemented in the WaveComBox toolbox. Boxes with solid lines are mandatory boxes, \textit{i.e.}, they consist in the building blocks of the modulation. On the other hand, boxes surrounded by dashed lines are optional. Some conventions regarding notations are introduced in the figures, including the number of information streams $S$, of transmit and receive signals, $N_T$ and $N_R$.

The \textbf{transmitter} consists of two key operations: generation of data symbols $\vect{d}$ and modulation of the transmitted signal $\vect{s}$. Additional operations can be included such as pre-equalization of the channel and/or the insertion of a preamble and pilot in the transmission frame.

The \textbf{channel} takes as an input the transmitted signal $\vect{s}$ and outputs the received signal $\vect{r}$. The channel can be viewed in a general sense as the transfer function between the discrete baseband samples at
the transmitter and the received baseband discrete samples at the receiver. In the ideal case, we have $\vect{r}=\vect{s}$. Otherwise, many impairments may be considered including additive noise and synchronization errors. Typical wireless effects are included such as multipath fading or mobility. The toolbox should be able to address optical effects as well such as optical fiber induced chromatic dispersion and laser phase noise.

The \textbf{receiver} takes as input the received signal $\vect{r}$ and aims at estimating the transmitted symbols $\hat{\vect{d}}$. A central block of the receiver is the demodulator. Other possible blocks implement synchronization, channel estimation and equalization and phase tracking.

In the WaveComBox toolbox, the waveform parameters are summarized in a structure that should be initialized at the beginning of each script. Examples of such parameters are the number of subcarriers, the number of data symbols, the constellation size, the number of transmit and receive antennas... This structure also contains some general parameters on the communication chain such as the signal-to-noise ratio or the velocity of the terminal. All parameters should not always be assigned to specific values depending on the scenario. For instance, velocity is only required if mobility is considered inducing a time-varying effect of the channel.

\section{First Example: FBMC-OQAM over AWGN}

The transmit signal $s[n]$ is obtained after FBMC-OQAM modulation of the purely real data symbols $d_{m,l}$, \textit{i.e.},
\begin{align*}
s[n]&=\sum_{m=0}^{2M-1}\sum_{l=0}^{2N_s -1} d_{m,l} g_{m,l}[n],
\end{align*}
where $g_{m,l}[n]=\jmath^{m+l}g[n-lM]e^{\jmath\frac{2\pi}{2M}m(n-lM-\frac{L_g-1}{2}) }$. Parameters $2M$, $2N_s$ and $L_g$ refer to the number of subcarriers, of real multicarrier symbols and to the length of the prototype filter $g[n]$. The received signal $r[n]$, affected by additive noise, is given by 
\begin{align*}
	r[n]=s[n] + w[n],
\end{align*}
where $w[n]$ are additive noise samples. The demodulated samples at subcarrier $m_0$ and multicarrier symbol $l_0$, are given by
\begin{align*}
z_{m_0,l_0}&=\sum_n r[n] g^*_{m_0,l_0}[n].
\end{align*}
Finally, the estimated symbols are obtained as 
\begin{align*}
	\hat{d}_{m_0,l_0}=\Re \left(z_{m_0,l_0}\right).
\end{align*}
In WaveComBox, this example can be simulated with the following 6 lines of code:

{\small
	\begin{verbatim} 
	% Initialization of FBMC-OQAM parameters
	Para = InitializeChainParameters('FBMC-OQAM'); 	
	
	% Transmitter
	d = GenerateData ( Para );
	s = Modulator( d, Para );	
	% AWGN channel
	r = Channel_AWGN( s, Para );		
	% Receiver
	z = Demodulator( r, Para );
	d_hat = real( z );
	\end{verbatim}}

This example is available in the WaveComBox toolbox under the name \texttt{BasicSISO.m}. The example is highly tunable as a function of different modulation parameters, channel effects and receiver architecture.

\section{Second Example: Optical Fiber FBMC-OQAM Chain}
The example \texttt{OpticalFiberChain.m} included in the toolbox allows the simulation of an optical FBMC-OQAM system based on the standard single-mode fiber (SSMF) limited by chromatic dispersion and laser phase noise \cite{7932847}. It implements the different signal processing operation blocks required to modulate the signal at the transmitter and to demodulate the signal at the receiver. The example can directly be used for experimental validation.

The first stage of the receiver relies on the preamble of the transmitted frame to perform synchronization and channel estimation. Synchronization relies on the repetitive pattern of the preamble to perform frame detection, symbol timing offset and carrier frequency offset (CFO) estimation according to \cite{thein2014analysis}. After synchronization is performed, the channel is estimated across frequency based on pilot symbols included in the preamble \cite{Rottenberg2015,Rottenberg2016}.

\begin{figure}[!t]
	\centering
	\resizebox{0.45\textwidth}{!}{%
		\Large
		\includegraphics[scale=1]{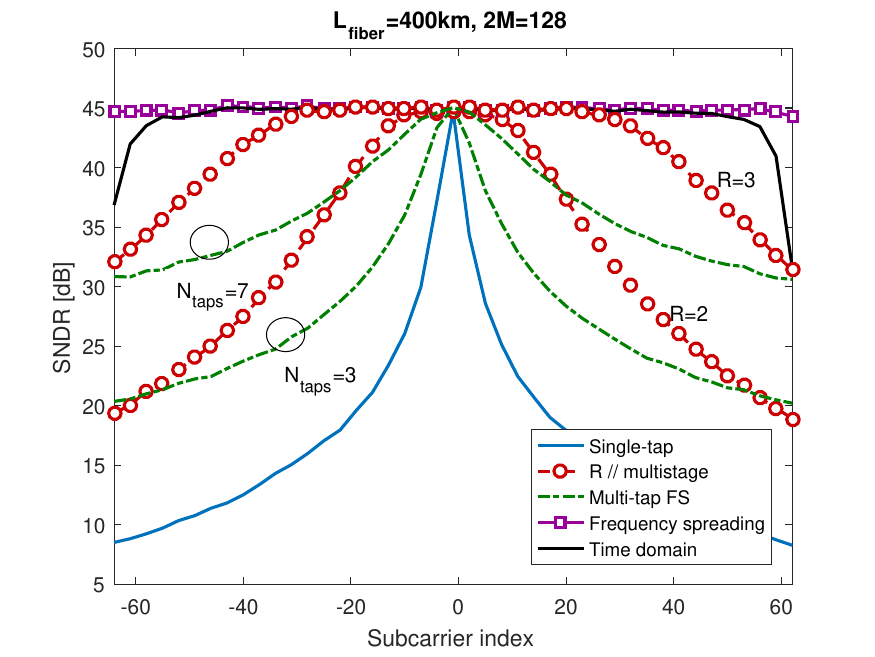}
	}
		\vspace{-1em}
	\caption{Chromatic dispersion compensation. Signal-to-noise-and-distortion ratio (SNDR) as a function of the subcarrier index when applying different chromatic dispersion compensation techniques. The fiber length is set to 400 km and the number of subcarriers is 128.}
	\label{fig:CD}
\end{figure}

Using estimated channel state information, the chromatic dispersion induced by the fiber can be compensated for. Different techniques are implemented and compared in the example \texttt{ChromaticDispersionCompensation.m} of the toolbox following the study presented in \cite{Rottenberg2017b} (see Fig.~\ref{fig:CD}). Single-tap equalization refers to the conventional FBMC-OQAM processing, similar to an OFDM system. The parallel equalization structure relies on a set of $R$ parallel analysis filterbanks architecture that implements a Taylor approximation of the ideal equalizer \cite{RottenbergTSPsingleTap}. The multi-tap equalizer consists of a fractionally spaced filter with memory $N_{taps}$ working at the subcarrier level. The frequency spreading structure, originally proposed by \cite{Bellanger2012}, uses an oversampled FFT at the receiver to perform a more accurate compensation. Finally, the time domain equalizer implements the conventional overlap-and-save algorithm prior to FMBC-OQAM demodulation.

Additionally, the finite linewidths of the transmit and receive lasers will lead to phase noise and hence a rotation of the received symbols in the complex plane. To fix this, phase tracking is implemented according to the methods presented in \cite{Rottenberg2017}.
Finally, after compensation of chromatic dispersion and phase noise, the symbols can be estimated by taking the real part of the demodulated samples.

\section{Third Example: MIMO FBMC-OQAM Equalization}

\begin{figure}[!t]
	\centering
	\resizebox{0.45\textwidth}{!}{%
		\Large
		\includegraphics[scale=1]{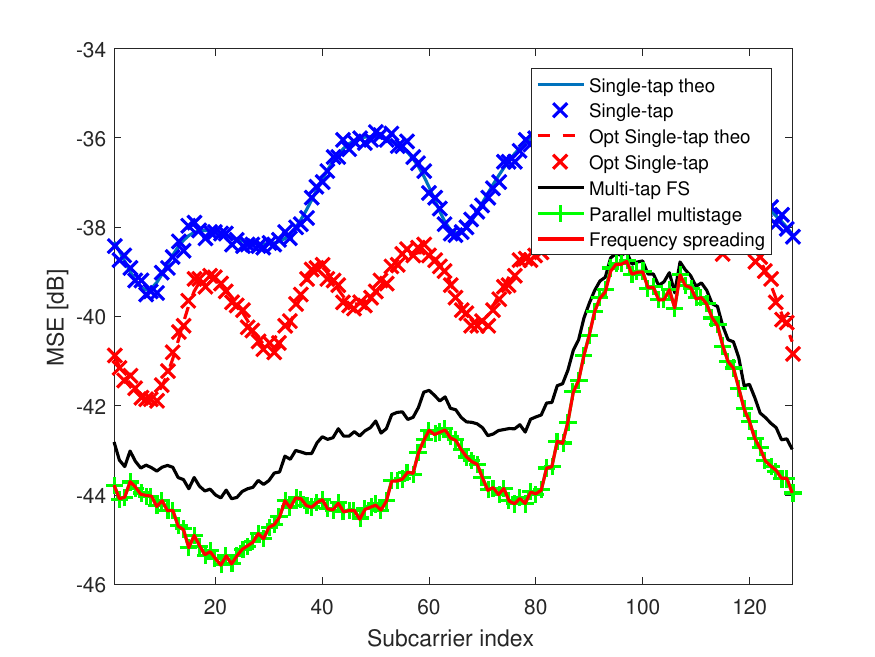}
	}
		\vspace{-1em}
	\caption{MIMO equalizer comparison.}
	\label{fig:MIMO_eq_compar}
\end{figure}

The example \texttt{MIMOEqualizerComparison.m} shows how to implement and compare the performance of a MIMO FBMC-OQAM system under a highly frequency selective wireless channel. Various equalization techniques are considered including the equalizers mentioned in the second example and additionally the low complexity single-tap equalizer of \cite{RottenbergTSPsingleTap}. Another example allows to evaluate the performance of a massive MIMO FBMC-OQAM system. The code allows to compare the simulated performance with the theoretical instantaneous and the asymptotic one derived in \cite{RottenbergTSPMassiveMIMO}.

\section{Conclusion}
\label{sec:conclusion}

In this article, we have described the general framework of the WaveComBox toolbox, being an open-source freely available piece of software. The toolbox relies on an abstract architecture that allows for simple simulation or experimental validation of communication chains using new waveforms. Moreover, it is highly documented and includes many useful examples. Some examples have a tutorial scope helping to apprehend the utilization of the toolbox. On the other hand, more comprehensive examples allow the full simulation of complete communication chains.


\vfill\pagebreak

\newpage
\bibliographystyle{IEEEbib}
\ninept

\bibliography{refs}

\begin{thebibliography}{10}

\bibitem{shafi20175g}
Mansoor Shafi, Andreas~F Molisch, Peter~J Smith, Thomas Haustein, Peiying Zhu,
  Prasan De~Silva, Fredrik Tufvesson, Anass Benjebbour, and Gerhard Wunder,
\newblock ``{5G: A Tutorial Overview of Standards, Trials, Challenges,
  Deployment, and Practice},''
\newblock {\em IEEE Journal on Selected Areas in Communications}, vol. 35, no.
  6, pp. 1201--1221, 2017.

\bibitem{wu2017overview}
Qingqing Wu, Geoffrey~Ye Li, Wen Chen, Derrick Wing~Kwan Ng, and Robert
  Schober,
\newblock ``{An overview of sustainable green 5G networks},''
\newblock {\em IEEE Wireless Communications}, vol. 24, no. 4, pp. 72--80, 2017.

\bibitem{li2006orthogonal}
Ye~Geoffrey Li and Gordon~L St\"{u}ber,
\newblock {\em {Orthogonal Frequency Division Multiplexing for Wireless
  Communications}},
\newblock Springer Science \& Business Media, 2006.

\bibitem{Sahin2014}
A.~Sahin, I.~Guvenc, and H.~Arslan,
\newblock ``{A Survey on Multicarrier Communications: Prototype Filters,
  Lattice Structures, and Implementation Aspects},''
\newblock {\em IEEE Communications Surveys Tutorials}, vol. 16, no. 3, pp.
  1312--1338, Third 2014.

\bibitem{6923528}
P.~Banelli, S.~Buzzi, G.~Colavolpe, A.~Modenini, F.~Rusek, and A.~Ugolini,
\newblock ``{Modulation Formats and Waveforms for 5G Networks: Who Will Be the
  Heir of OFDM?: An overview of alternative modulation schemes for improved
  spectral efficienc}y,''
\newblock {\em IEEE Signal Processing Magazine}, vol. 31, no. 6, pp. 80--93,
  Nov 2014.

\bibitem{horlin2013dual}
Fran{\c{c}}ois Horlin, Jessica Fickers, Philippe Emplit, Andr{\'e} Bourdoux,
  and J{\'e}rome Louveaux,
\newblock ``{Dual-polarization OFDM-OQAM for communications over optical fibers
  with coherent detection},''
\newblock {\em Optics express}, vol. 21, no. 5, pp. 6409--6421, 2013.

\bibitem{7932847}
T.~H. Nguyen, F.~Rottenberg, S.~P. Gorza, J.~Louveaux, and F.~Horlin,
\newblock ``{Efficient Chromatic Dispersion Compensation and Carrier Phase
  Tracking for Optical Fiber FBMC/OQAM Systems},''
\newblock {\em Journal of Lightwave Technology}, vol. 35, no. 14, pp.
  2909--2916, July 2017.

\bibitem{Rottenberg18}
F.~Rottenberg, P.~T. Dat, T.~H. Nguyen, A.~Kanno, F.~Horlin, J.~Louveaux, and
  N.~Yamamoto,
\newblock ``{2 x 2 MIMO FBMC-OQAM Signal Transmission Over a Seamless
  Fiber--Wireless System in the W-band},''
\newblock {\em IEEE Photonics Journal}, vol. 10, no. 2, pp. 1--14, April 2018.

\bibitem{rottenberg2018fbmc}
Fran{\c{c}}ois Rottenberg,
\newblock ``{FBMC-OQAM transceivers for wireless and optical fiber
  communications},''
\newblock 2018.

\bibitem{lin2016experimental}
Bangjiang Lin, Xuan Tang, Zabih Ghassemlooy, Xi~Fang, Chun Lin, Yiwei Li, and
  Shihao Zhang,
\newblock ``{Experimental demonstration of OFDM/OQAM transmission for visible
  light communications},''
\newblock {\em IEEE Photonics Journal}, vol. 8, no. 5, pp. 1--10, 2016.

\bibitem{van2017performance}
Mathieu Van~Eeckhaute, Andr{\'e} Bourdoux, Philippe De~Doncker, and
  Fran{\c{c}}ois Horlin,
\newblock ``{Performance of emerging multi-carrier waveforms for 5G
  asynchronous communications},''
\newblock {\em EURASIP Journal on wireless communications and networking}, vol.
  2017, no. 1, pp. 29, 2017.

\bibitem{gerzaguet20175g}
Robin Gerzaguet, Nikolaos Bartzoudis, Leonardo~Gomes Baltar, Vincent Berg,
  Jean-Baptiste Dor{\'e}, Dimitri Kt{\'e}nas, Oriol Font-Bach, Xavier Mestre,
  Miquel Payar{\'o}, Michael F{\"a}rber, et~al.,
\newblock ``{The 5G candidate waveform race: a comparison of complexity and
  performance},''
\newblock {\em EURASIP Journal on Wireless Communications and Networking}, vol.
  2017, no. 1, pp. 13, 2017.

\bibitem{Zayani2016}
R.~{Zayani}, Y.~{Medjahdi}, H.~{Shaiek}, and D.~{Roviras},
\newblock ``{WOLA-OFDM: A Potential Candidate for Asynchronous 5G},''
\newblock in {\em 2016 IEEE Globecom Workshops (GC Wkshps)}, Dec 2016, pp.
  1--5.

\bibitem{farhang2011ofdm}
Behrouz Farhang-Boroujeny,
\newblock ``{OFDM versus filter bank multicarrier},''
\newblock vol. 28, no. 3, pp. 92--112, May 2011.

\bibitem{Vakilian2013}
V.~Vakilian, T.~Wild, F.~Schaich, S.~ten Brink, and J.~F. Frigon,
\newblock ``{Universal-filtered multi-carrier technique for wireless systems
  beyond LTE},''
\newblock in {\em 2013 IEEE Globecom Workshops (GC Wkshps)}, Dec 2013, pp.
  223--228.

\bibitem{Abdoli2015}
J.~Abdoli, M.~Jia, and J.~Ma,
\newblock ``{Filtered OFDM: A new waveform for future wireless systems},''
\newblock in {\em 2015 IEEE 16th International Workshop on Signal Processing
  Advances in Wireless Communications (SPAWC)}, June 2015, pp. 66--70.

\bibitem{thein2014analysis}
Christoph Thein, Malte Schellmann, and J{\"u}rgen Peissig,
\newblock ``{Analysis of frequency domain frame detection and synchronization
  in OQAM-OFDM systems},''
\newblock {\em EURASIP Journal on Advances in Signal Processing}, vol. 2014,
  no. 1, pp. 83, 2014.

\bibitem{Rottenberg2015}
F.~{Rottenberg}, Y.~{Medjahdi}, E.~{Kofidis}, and J.~{Louveaux},
\newblock ``{Preamble-based channel estimation in asynchronous FBMC-OQAM
  distributed MIMO systems},''
\newblock in {\em 2015 International Symposium on Wireless Communication
  Systems (ISWCS)}, Aug 2015, pp. 566--570.

\bibitem{Rottenberg2016}
F.~{Rottenberg}, F.~{Horlin}, E.~{Kofidis}, and J.~{Louveaux},
\newblock ``{Generalized optimal pilot allocation for channel estimation in
  multicarrier systems},''
\newblock in {\em 2016 IEEE 17th International Workshop on Signal Processing
  Advances in Wireless Communications (SPAWC)}, July 2016, pp. 1--5.

\bibitem{Rottenberg2017b}
F.~{Rottenberg}, T.~{Nguyen}, S.~{Gorza}, F.~{Horlin}, and J.~{Louveaux},
\newblock ``{Advanced Chromatic Dispersion Compensation in Optical Fiber
  FBMC-OQAM Systems},''
\newblock {\em IEEE Photonics Journal}, vol. 9, no. 6, pp. 1--10, Dec 2017.

\bibitem{Rottenberg2017}
F.~{Rottenberg}, T.~{Nguyen}, S.~{Gorza}, F.~{Horlin}, and J.~{Louveaux},
\newblock ``{ML and MAP phase noise estimators for optical fiber FBMC-OQAM
  systems},''
\newblock in {\em 2017 IEEE International Conference on Communications (ICC)},
  May 2017, pp. 1--6.

\bibitem{Bellanger2012}
M.~{Bellanger},
\newblock ``{FS-FBMC: An alternative scheme for filter bank based multicarrier
  transmission},''
\newblock in {\em 2012 5th International Symposium on Communications, Control
  and Signal Processing}, May 2012, pp. 1--4.

\bibitem{RottenbergTSPsingleTap}
F.~{Rottenberg}, X.~{Mestre}, F.~{Horlin}, and J.~{Louveaux},
\newblock ``{Single-Tap Precoders and Decoders for Multiuser MIMO FBMC-OQAM
  Under Strong Channel Frequency Selectivity},''
\newblock {\em IEEE Transactions on Signal Processing}, vol. 65, no. 3, pp.
  587--600, Feb 2017.

\bibitem{ihalainen2006channel}
Tero Ihalainen, Tobias~Hidalgo Stitz, Mika Rinne, and Markku Renfors,
\newblock ``{Channel equalization in filter bank based multicarrier modulation
  for wireless communications},''
\newblock {\em EURASIP Journal on Advances in Signal Processing}, vol. 2007,
  no. 1, pp. 049389, 2006.

\bibitem{RottenbergTSPParallel}
F.~{Rottenberg}, X.~{Mestre}, D.~{Petrov}, F.~{Horlin}, and J.~{Louveaux},
\newblock ``{Parallel Equalization Structure for MIMO FBMC-OQAM Systems Under
  Strong Time and Frequency Selectivity},''
\newblock {\em IEEE Transactions on Signal Processing}, vol. 65, no. 17, pp.
  4454--4467, Sep. 2017.

\bibitem{RottenbergTSPMassiveMIMO}
F.~{Rottenberg}, X.~{Mestre}, F.~{Horlin}, and J.~{Louveaux},
\newblock ``{Performance Analysis of Linear Receivers for Uplink Massive MIMO
  FBMC-OQAM Systems},''
\newblock {\em IEEE Transactions on Signal Processing}, vol. 66, no. 3, pp.
  830--842, Feb 2018.

\end{thebibliography}

\end{document}